\documentclass[superscriptaddress,twocolumn,prb,aps,10pt]{revtex4-1}

\usepackage{amsmath}
\usepackage{amsfonts}
\usepackage{amssymb}
\usepackage{graphicx}

\usepackage{xcolor}
\definecolor{linkblue}{RGB}{6,69,173}
\usepackage{hyperref}
\hypersetup{
	pdffitwindow=true,     													
    pdfstartview={FitV},    												
    pdftitle={Cosacchi_2021_Deterministic},    
    pdfauthor={M. Cosacchi et. al.},  											
    pdfsubject={Research Article},   											
    pdfkeywords={single photons, storage, quantum emitters, optical cavities, quantum optics}, 			
    linktoc=page,
	colorlinks=true,
	allcolors=linkblue,
}
\usepackage{footnote}
\usepackage{textcomp}

\usepackage{upgreek}
\usepackage{bm}
\usepackage{endnotes}
\let\footnote=\endnote

\newcommand{\OP}[2]{\vert #1\rangle\langle #2\vert}
\newcommand{\BRA}[1]{\langle #1\vert}
\newcommand{\KET}[1]{\vert #1\rangle}

\newcommand{\LB}[2]{\mathcal{L}_{#1,#2}}

\DeclareUnicodeCharacter{2009}{\,}

\renewcommand{\t}[1]{\textrm{#1}}
\newcommand{\nn}{\nonumber\\}

\newcommand{\q}{\bm{q}}

\renewcommand{\a}{\alpha}

\newcommand{\g}{\gamma}

\renewcommand{\r}{\rho}
\newcommand{\s}{\sigma}

\newcommand{\w}{\omega}

\renewcommand{\L}{\mathcal{L}}

\newcommand{\+}{^\dagger}
\renewcommand{\>}{\rangle}

\newcommand{\G}{\vert G\>}
\newcommand{\X}{\vert X\>}

\newcommand{\D}{\vert D\>}

\newcommand{\GG}{\vert G\>\<G\vert}
\newcommand{\XX}{\vert X\>\<X\vert}

\newcommand{\DD}{\vert D\>\<D\vert}

\newcommand{\XG}{\vert X\>\<G\vert}
\newcommand{\GX}{\vert G\>\<X\vert}

\newcommand{\DX}{\vert D\>\<X\vert}
\newcommand{\XD}{\vert X\>\<D\vert}

\newcommand{\GD}{\vert G\>\<D\vert}

\newcommand{\ddt}{\frac{\partial}{\partial t}}

\newcommand{\deff}{\delta_\text{eff}}
\newcommand{\hs}{;\hspace{1cm}}

\begin{document}


\title{Deterministic photon storage and readout in a semimagnetic quantum~dot--cavity~system doped with a single Mn ion}

\author{Michael Cosacchi*}
\affiliation{Lehrstuhl f{\"u}r Theoretische Physik III, Universit{\"a}t Bayreuth, Universit{\"a}tsstraße 30, 95447 Bayreuth, Germany}
\email[]{michael.cosacchi@uni-bayreuth.de}
\author{Tim Seidelmann}
\affiliation{Lehrstuhl f{\"u}r Theoretische Physik III, Universit{\"a}t Bayreuth, Universit{\"a}tsstraße 30, 95447 Bayreuth, Germany}
\author{Adam Mielnik-Pyszczorski}
\affiliation{Lehrstuhl f{\"u}r Theoretische Physik III, Universit{\"a}t Bayreuth, Universit{\"a}tsstraße 30, 95447 Bayreuth, Germany}
\affiliation{Department of Theoretical Physics, Wroc{\l}aw University of Science and Technology, 50-370 Wroc{\l}aw, Poland}
\author{Miriam Neumann}
\affiliation{Institut f{\"u}r Festk{\"o}rpertheorie, Universit{\"a}t M{\"u}nster, Wilhelm-Klemm-Straße 10, 48149 M{\"u}nster, Germany}
\author{Thomas K. Bracht}
\affiliation{Institut f{\"u}r Festk{\"o}rpertheorie, Universit{\"a}t M{\"u}nster, Wilhelm-Klemm-Straße 10, 48149 M{\"u}nster, Germany}
\author{Moritz Cygorek}
\affiliation{Institute of Photonics and Quantum Sciences, Heriot-Watt University, Edinburgh EH14 4AS, United Kingdom}
\author{Alexei Vagov}
\affiliation{Lehrstuhl f{\"u}r Theoretische Physik III, Universit{\"a}t Bayreuth, Universit{\"a}tsstraße 30, 95447 Bayreuth, Germany}
\affiliation{ITMO University, Kronverksky Pr. 49, St. Petersburg, 197101, Russia}
\author{Doris E. Reiter}
\affiliation{Institut f{\"u}r Festk{\"o}rpertheorie, Universit{\"a}t M{\"u}nster, Wilhelm-Klemm-Straße 10, 48149 M{\"u}nster, Germany}
\author{Vollrath M. Axt}\affiliation{Lehrstuhl f{\"u}r Theoretische Physik III, Universit{\"a}t Bayreuth, Universit{\"a}tsstraße 30, 95447 Bayreuth, Germany}

\begin{abstract}
Light trapping is a crucial mechanism for synchronization in optical communication.
Especially on the level of single photons, control of the exact emission time is desirable.
In this paper, we theoretically propose a single-photon buffering device composed of a quantum dot doped with a single Mn atom in a cavity.
We present a method to detain a single cavity photon as an excitation of the dot.
The storage scheme is based on bright to dark exciton conversion performed with an off-resonant external optical field and mediated via a spin-flip with the magnetic ion.
The induced Stark shift brings both exciton states to resonance and results in an excitation transfer to the optically inactive one.
The stored photon can be read out on demand in the same manner by repopulating the bright state, which has a short lifetime.
Our results indicate the possibility to suspend a photon for almost two orders of magnitude longer than the lifetime of the bright exciton.
\end{abstract}

\maketitle

\section{Introduction}
\label{sec:introduction}


Self-assembled quantum dots (QDs) are optically active and allow the control of electronic states with light.\cite{ramsay2010ReviewCoherentOptical, greilich2006OpticalControlSpina, kroner2008OpticalDetectionSingleElectrona, dutt2005StimulatedSpontaneousOpticala}
In return, they can serve as photon sources, which makes them attractive for quantum communication devices.\cite{greve2013UltrafastOpticalControla, gao2015CoherentManipulationMeasurementa}
A QD--cavity system greatly increases light emission efficiency due to the Purcell effect \cite{Purcell1946} and has a favored direction of emission in contrast to a standalone QD, providing easier in- and outcoupling.
While QDs in cavities are a suitable platform for quantum information processing devices,\cite{hennessy2007QuantumNatureStrongly} the realization requires the synchronization of signals,\cite{takesue2013OnchipCoupledResonator} for which a photon buffer is desirable.

In all-optical systems, buffers were realized with fibers and waveguides.\cite{takesue2013OnchipCoupledResonator, liu2016WavelengthTunableOptical, clemmen2018AllopticallyTunableBuffer, lai2018UltrawidebandStructuralSlow}
Another proposed realization of an optical memory cell is a three-level $\Lambda$ system, which gives the possibility to store information in a dark state. \cite{liu2001ObservationCoherentOptical}
Extremely long light storage was achieved in atomic systems, where the slow light effect is commonly based on electromagnetically induced transparency (EIT).
Adequate coupling in the $\Lambda$ system highly reduces the group velocity of light and results in slow propagation of a beam through atoms or even reversible trapping of light in atomic excitations.\cite{hau1999LightSpeedReduction, liu2001ObservationCoherentOptical, dudin2013LightStorageTime, katz2018LightStorageOne}
Recent experiments were performed even on the single-photon level,\cite{deriedmatten2008SolidstateLightMatter, rakher2013ProspectsStorageRetrieval, saglamyurek2019SinglephotonlevelLightStorage} raising hopes for use in quantum communication.
Atomic systems were also used to store time-entangled solitons in a cavity, representing a step towards multiplexed quantum communication. \cite{welakuh2017StorageRetrievalTimeentangled}

In solid-state systems, photons may be absorbed and stored as excitons.
However, the typical lifetime of a bright exciton is short (typically a few hundred ps up to one ns). Therefore, a separated electron-hole pair, the indirect exciton, was used in coupled nanostructures to extend the storage time.\cite{lundstrom1999ExcitonStorageSemiconductor, song-bao2003PhotonStorageOpticalMemory, winbow2007PhotonStorageNanosecond, winbow2008PhotonStorageSubnanosecond, high2008ControlExcitonFluxes, fischer2009ExcitonStorageNanoscale, simonin2014ElectricMagneticField, climente2014ExcitonStorageTypeII, tseng2015UsingDarkStates}
On the other hand, in a single QD--cavity system, the lifetime of an exciton may be increased by the Stark shift, which decouples the exciton from the cavity mode.\cite{johne2011SinglephotonAbsorptionDynamic}
A more attractive direction for storing excitations in a QD is to use a dark state, which lives for at least an order of magnitude longer than the bright one.
For a long time, dark excitons were beyond much interest as they are not optically active and hence not directly accessible.

Recent progress allows for indirectly accessing the dark exciton with light\cite{poem2010AccessingDarkExciton, zielinski2020FineStructureDark} or other complexes, \cite{mcfarlane2009GigahertzBandwidthElectrical} but all-optical control of the dark state using an intermediate biexciton state has also been proposed for the use as a long-lived qubit.\cite{poem2010AccessingDarkExciton, schwartz2015DeterministicCoherentWriting, schwartz2015DeterministicWritingControl, heindel2017AccessingDarkExciton, schmidgall2017CoherentControlDark}
Another possibility is the coupling of bright and dark states using micromechanical resonators, making the dark state addressable by light.\cite{ohta2018DynamicControlCoupling}
Still, it is much easier to excite the bright state.
Hence, a method to realize the bright-to-dark conversion for excitation storage was already proposed for colloidal systems.\cite{kraus2007RoomTemperatureExcitonStorage}
Yet, they cannot be easily integrated on-chip.
A dark state was also used as a microsecond valley polarization memory in transition metal dichalcogenides.\cite{jiang2018MicrosecondDarkexcitonValley}
More recently, a controllable occupation transfer between bright and dark excitons in a QD--cavity system was suggested.\cite{Neumann2021}

Our method facilitates QDs with a single magnetic dopant, which can be deterministically fabricated for several years,\cite{Kobak2014} with dopand atoms like Manganese (Mn),\cite{Besombes2004,Goryca2010} Cromium,\cite{Lafuente-Sampietro2017} Iron,\cite{Smolenski2016} or Cobalt.\cite{Kobak2018}
Interestingly, the spin state of the dopant can be changed via optical control.\cite{Reiter2009,Goryca2009,LeGall2009,Reiter2013,Lafuente-Sampietro2016}
Here, we choose a Mn doped CdTe/ZnTe QD in a microcavity.\cite{Pacuski2014}
The exchange coupling between the Mn and the electron spin enables a coupling between bright and dark excitons in the quantum dot under the simultaneous flip of the Mn spin.\cite{Reiter2009,Reiter2012}
Given a spin state, the states can be interpreted as a $\Lambda$-type three-level system, as specified in the next section.

The buffering scheme relies on storing the photon in the dark state.
After the photon is converted to the bright exciton state, an AC-Stark pulse is utilized to facilitate the conversion of the bright into a dark exciton.
We stress that the Stark pulse is the only external pulse which is used in the buffering scheme. Because the coupling between bright and dark excitons in this system is enabled by the exhange interaction with the Mn dopant, no external magnetic field needs to be applied as in other studies.\cite{Luker2015,Luker2017b,Neumann2021}.
This is a significant advantage because it gives a possibility to integrate magnetically doped QDs into compact on-chip devices.




\begin{table*}
\begin{center}
\caption{Parameters used for the simulations.}
  \begin{tabular}{ l  c  c	r }
  
    \hline
    \hline
    Electron-Mn coupling				& $J_e$	[meV nm$^3$]		& $-15$		& \cite{Furdyna1988}\\ 
    Hole-Mn coupling					& $J_h$	[meV nm$^3$]		& $60$		& \cite{Furdyna1988}\\
    Intrinsic dark-bright splitting				& $\delta_{\t{XD}}$ [meV]	& $0.95$	& \cite{Besombes2002}\\
    Mn g-factor 						& $g_{\t{Mn}}$				& $2.0075$	& \cite{Leger2005}\\
    Electron g-factor					& $g_{\t{e}}$				& $-1.5$	& \cite{Besombes2004}\\
    QD--cavity coupling 				& $\hbar g$ [meV]			& $0.1$	& \cite{Jakubczyk2012}\\
    Cavity loss rate 					& $\kappa$ [ns$^{-1}$]		& $8.5$	& \cite{Schneider2016}\\
    Radiative decay rate of $\X$		& $\g_{\t{X}}$ [ns$^{-1}$]	& $2.4$	& \cite{Jakubczyk2012}\\
    Residual decay rate of $\D$			& $\g_{\t{D}}$ [ns$^{-1}$]	& $0.01$	& \cite{McFarlane2009}\\
    Electron deformation potential		& $D_e$ [eV]				& $-5$ 	& \cite{Besombes2001}\\
    Hole deformation potential			& $D_h$ [eV]				& $1$ 	& \cite{Besombes2001}\\
    Density								& $\r_D$ [kg m$^{-3}$]		& $5510$	& \cite{Besombes2001}\\
    Sound velocity						& $c_s$ [m s$^{-1}$]				& $4000$	& \cite{Besombes2001}\\
    Electron-to-hole confinement ratio	& $a_e/a_h$					& $1.38$	& \cite{ChinaRare2021,Luker2017}\\
    Electron confinement radius			& $a_e$ [nm]				& $3.0$	& \cite{Cygorek2017,Quilter2015,Bounouar2015}\\    
    \hline
    \hline
  \end{tabular}
  \label{tab:par}
\end{center}
\end{table*}

\section{Mn-doped quantum dot system}
\label{sec:Mn}

\begin{figure}[t]
\centering
\includegraphics{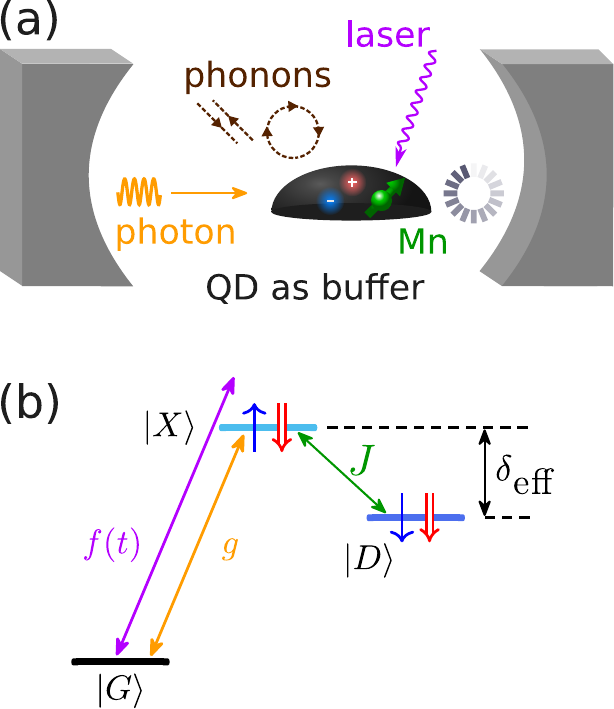}
\caption{(a) Sketch of the QD--cavity system.
The Mn ion provides the magnetic field necessary to facilitate the storage of a single cavity photon.
An off-resonant external laser controls both the storage and the release time and thus the entire buffering procedure deterministically.
(b) The $\Lambda$-type three-level model of the QD.
The spin configuration of the two exciton states is symbolized by arrows.
}
\label{fig:sketch}
\end{figure}

We consider a self-assembled CdTe quantum dot (QD) doped by a single Mn ion inside a ZnTe micropillar cavity.
Due to the strong spatial confinement of the carriers in the QD, only the lowest conduction band state and the uppermost valence band state need to be considered, namely electrons in the $s$-like conduction band and holes in the $p$-like heavy-hole band.

Excitons form as pairs of conduction band electrons and valence band heavy holes.
Having a spin component of $S_z^{\t{h}}=\pm\frac{3}{2}$, heavy holes can form two types of excitons with the spin-$\frac{1}{2}$ electrons:
the optically active \textit{bright} states with a circular polarization of $\pm1$ and the dipole-\textit{dark} states with $\pm2$.
For typical fine-structure splittings of a few tens of $\upmu$eV\cite{Bounouar2018} between the two bright exciton states of opposite circular polarization, only excitons of one polarization need to be considered, if the external driving has a defined circular polarization.\cite{Luker2015,Luker2017a,Cosacchi2020b}

Doping such a QD system with a single Mn ion, which has a spin of $\frac{5}{2}$, introduces an additional state space, namely the six possible orientations of its spin.
The Mn spin interacts with electrons and holes via the exchange interaction\cite{Besombes2004,Leger2005,Leger2007,Besombes2008,
Reiter2009,Reiter2009b}
\begin{align}
H_{\t{ex}}=j_{\t{e}}\bm{M}\cdot\bm{S}^{\t{e}}
+j_{\t{h}}\bm{M}\cdot\bm{S}^{\t{h}}\, ,
\end{align}
where $\bm{M}$ denotes the spin of the Mn ion.
$\bm{S}^{\t{e}}$ ($\bm{S}^{\t{h}}$) is the operator of the electron (hole) spin in the QD.
$j_{\t{e/h}}=J_{\t{e/h}}\left|\Psi_0^{\t{e/h}}(\bm{r}_{\t{Mn}})\right|^2$ are composed of the coupling constants $J_{\t{e/h}}$ between the electon/hole and the Mn spin (cf., \textbf{Table~\ref{tab:par}}) and the carrier ground state wave function $\Psi_0$ at the position $\bm{r}_{\t{Mn}}$ of the Mn atom.
Modelling the QD with a hard wall cubic potential \cite{Fernandez-Rossier2006,Reiter2009b} with in-plane widths of $6\,$nm and a height of $2\,$nm, the coupling strengths $j_{\t{e/h}}$ depends on the position of the Mn atom.

For a more intuitive understanding of the exchange Hamiltonian, it can be rewritten as
\begin{align}
\label{eq:H_ex}
H_{\t{ex}}=j_{\t{e}}M_z S_z^{\t{e}}
+\frac{j_{\t{e}}}{2}\left(M_+ S_-^{\t{e}} + M_- S_+^{\t{e}}\right)\nn
+j_{\t{h}}M_z S_z^{\t{h}}
+\frac{j_{\t{h}}}{2}\left(M_+ S_-^{\t{h}} + M_- S_+^{\t{h}}\right)
\end{align} 
with $M_\pm:=M_x\pm iM_y$ and $S_\pm^{\t{e/h}}:=S_x^{\t{e/h}}\pm iS_y^{\t{e/h}}$.
The Ising terms\footnote{The terms proportional to $M_z$ in Equation~\eqref{eq:H_ex}} arising from the $z$-component of the interaction lead to energy shifts of the exciton states with different spin configuration.
These contributions lead to the characteristic splitting of the exciton line into six lines even at zero magnetic field.\cite{Besombes2004,Besombes2008,Goryca2010,Kobak2014}
The electron flip-flop term\footnote{The terms proportional to $j_e M_\pm$ in Equation~\eqref{eq:H_ex}} on the other hand results in a coupling between the excitonic bright state with total spin $\pm1$ and the excitonic dark state with $\pm2$ via simultaneous spin flip.
While usually the flip-flop term is much weaker than the energetic splitting, for an applied magnetic field in Faraday configuration, this coupling is seen as anti-crossing in the optical spectrum at a field of several Tesla \cite{Slavcheva2010,Besombes2005}.
Note that the flip-flop term regarding the hole\footnote{The terms proportional to $j_h M_\pm$ in Equation~\eqref{eq:H_ex}} can be neglected since the hole spin is pinned in a pure heavy-hole system.\cite{Reiter2009}

Assuming the Mn spin to be initially prepared in the state $M_z=-\frac{5}{2}$, we can reduce our system to a three-level system.
This preparation can be achieved in numerous ways: by thermal occupation at low temperatures since by applying a magnetic field, it becomes the energetically lowest state \cite{Reiter2009};
or by an all-optical protocol \cite{LeGall2010}, thus avoiding the necessity of an additional external magnetic field.
Then, the three states are:
the ground state without an electronic excitation $\G:=\vert 0,-\frac{5}{2}\rangle$, the bright exciton $\X:=\vert -1,-\frac{5}{2}\rangle$, and the dark exciton $\D:=\vert -2,-\frac{3}{2}\rangle$.
Here, the first entry denotes the projection of the total spin of the electronic excitation and the second one the Mn spin orientation.
Note that a circular polarization of the external laser of $-1$ is assumed, from which the sign of the bright exciton spin follows.

\section{Model of the $\Lambda$-type three-level system}
\label{sec:model}


In the basis of the three states $\G$, $\X$, and $\D$ the Hamiltonian reads as follows:
\begin{align}
H=H_{\t{QD}} + H_{\t{flip}} + H_{\t{driv}}(t) + H_{\t{C}} + H_{\t{Ph}}\, ,
\end{align}
consisting of the QD part $H_{\t{QD}}$ and the flip-flop term $H_{\t{flip}}$ as introduced in Equation~\eqref{eq:H_ex}.
In addition, we account for the driving of the system with an external laser pulse $H_{\t{driv}}(t)$, the coupling to a single-mode cavity $H_{\t{C}}$, and the coupling to longitudinal acoustic (LA) phonons $H_{\t{Ph}}$.
A sketch of the system and its level structure is shown in \textbf{Figure~\ref{fig:sketch}}.

The QD part is composed of
\begin{align}
H_{\t{QD}}=\hbar\w_{\t{X}}\XX + \left(\hbar\w_{\t{X}}-\delta_{\t{eff}}\right)\DD\, ,
\end{align}
where the energy of the ground state is set to zero, the bright exciton has the energy $\hbar\w_{\t{X}}$, and the effective dark-bright splitting is $\delta_{\t{eff}}$.
Three contributions enter the latter quantity:
the intrinsic splitting $\delta_{\t{XD}}$ due to the electron-hole exchange interaction and the splitting arising from the Ising terms in Eq.~\eqref{eq:H_ex}
\begin{align}
\label{eq:delta_eff}
\delta_{\t{eff}}=\delta_{\t{XD}}-2 j_{\t{e}}+\frac{3}{2}j_{\t{h}}+\left(g_{\t{Mn}}-g_{\t{e}}\right)\mu_{\t{B}}B_z\, .
\end{align}
The third contribution is a Zeeman splitting due to an external magnetic field in Faraday configuration $\bm{B}=B_z\bm{e}_z$.
$g_{\t{Mn}}$ and $g_{\t{e}}$ in Eq.~\eqref{eq:delta_eff} denote the Mn and the electron $g$-factors (cf., Table~\ref{tab:par}), respectively, and $\mu_{\t{B}}$ is the Bohr magneton.
 
One arm of our $\Lambda$-type system is coupled by the flip-flop term
\begin{align}
H_{\t{flip}}=-\frac{1}{2}J\left(\XD+\DX\right)\, .
\end{align}
The interaction strength results from calculating the corresponding matrix elements in the three-level basis as $J=-\sqrt{5}j_{\t{e}}$.
We assume the position of the Mn atom to be $30\,\%$ away from the QD edge in both $x$ and $y$ direction and $13\,\%$ in $z$ direction.
This results in a coupling strength of $J=0.25\,$meV and an effective dark-bright splitting of $\delta_{\t{eff}}=1.85\,$meV in the field-free case $B_z=0$.
This value can be interpreted as the Mn spin providing an effective magnetic field for the excitons with a strength of roughly $3\,$T.

The other arm of the $\Lambda$-type system, i.e., the ground to bright exciton state transition is driven by an external laser classically described by the function $f(t)=f_{\t{ACS}}(t)e^{-i\w_{\t{ACS}}t}$ with the real envelope function $f_{\t{ACS}}(t)$ and the off-resonant AC-Stark frequency $\w_{\t{ACS}}$ [cf., Figure~\ref{fig:sketch}(b)].
Although the AC-Stark pulse is off-resonant, the parameters are chosen such that the conditions for the usual dipole and rotating wave approximations still hold and the corresponding coupling can be written as:\cite{Jaynes1963,Shore1993}
\begin{align}
H_{\t{driv}}(t)=-\frac{\hbar}{2}\left(f^*(t)\GX+f(t)\XG\right)\, .
\end{align}
The coupling to the single-mode cavity with strength $g$ [cf., Figure~\ref{fig:sketch}(b)] is described by a Jaynes-Cummings model
\begin{align}
H_{\t{C}}=\hbar\w_{\t{C}} a\+ a + \hbar g \left(a\XG + a\+ \GX\right)\, ,
\end{align}
where $a$ ($a\+$) is the annihilation (creation) operator for a photon at the cavity frequency $\w_{\t{C}}$, which is assumed to be on resonance with the bright state $\w_{\t{X}}$.

To model the decoherence in the QD, we consider that the QD is coupled to an environment of LA phonons in the bulk material [cf., Figure~\ref{fig:sketch}(a)]\cite{Besombes2001,Borri2001,
Krummheuer2002,Axt2005,Reiter2014,Reiter2019}
\begin{align}
\label{eq:H_Ph}
H_{\t{Ph}}=\,&\hbar\sum_{\q} \w_{\q} b_{\q}\+ b_{\q}\nn
&+\hbar\sum_{\q} \left(\g_{\q}b_{\q}\+ +\g_{\q}^* b_{\q}\right)\left(\XX+\DD\right)\, .
\end{align}
$b_{\q}$ ($b_{\q}\+$) annihilates (creates) a phonon in the mode $\q$ with the frequency $\w_{\q}$.
Both exciton states are assumed to couple to the environment with the same strength $\g_{\q}$.
The role of phonons in QD--cavity systems is typically considered to be detrimental to the preparation of photonic quantum states, e.g., single photons\cite{Michler2000,Santori2001,Santori2002,He2013,
Wei2014Det,Ding2016,Somaschi2016,
Schweickert2018,Hanschke2018} or entangled photon pairs.\cite{Akopian2006,Stevenson2006a,Hafenbrak2007,Dousse2010,
delvalle2013dis,Mueller2014,Orieux2017}
Nonetheless, in specific situations a phonon enhancement of the single-photon purity is found.\cite{Cosacchi2019,Thomas2021}
Also, a boost in the entanglement of two photons has been predicted to be a result of the phonon interaction.\cite{Seidelmann2019}

Furthermore, we account for cavity losses ($\L_{a,\kappa}$) as well as radiative decay of the bright exciton ($\L_{\GX,\gamma_{\t{X}}}$) and losses of the dark exciton ($\L_{\GD,\gamma_{\t{D}}}$) using Lindblad superoperators acting on the density matrix $\rho$ as
\begin{align}
\mathcal{L}_{O,\Gamma}\rho=\Gamma\left(O\rho O\+ -\frac{1}{2}\left\lbrace\rho,O\+ O\right\rbrace_+\right)\, ,
\end{align}
where $\left\lbrace A,B\right\rbrace_+$ is the anti-commutator of operators $A$ and $B$.
These superoperators describe phenomenologically loss processes with rate $\Gamma$ on a dissipation channel $O$. 

We use an interaction picture representation of this Hamiltonian for the numerics as well as the physical discussion, in order to eliminate fast oscillating terms in the dynamics resulting from transition energies in the eV range.
The noninteracting Hamiltonian used for this transform is\cite{Neumann2021}
\begin{align}
H_0=\,&-\hbar\Delta\w_{\t{AX}} \GG +\hbar\w_{\t{X}}\left(\XX+\DD\right)\nn
&+\hbar\w_{\t{ACS}}a\+ a\, .
\end{align}
Here, the detuning between the laser and bright exciton frequencies $\Delta\w_{\t{AX}}:=\w_{\t{ACS}}-\w_{\t{X}}$ has been introduced.
Then, the transformed Hamiltonian is $H_I=U\+ (H-H_0) U$ with $U=\exp{[-(i/\hbar) H_0 t]}$.

We choose parameters from the experimental literature, in order to perform simulations as realistic as possible.
The values are given in Table~\ref{tab:par} together with corresponding references.

The dynamics is obtained as the solution of the Liouville-von Neumann equation
\begin{align}
\label{eq:Liouville-von_Neumann}
\frac{\partial}{\partial t} \r =\,&
-\frac{i}{\hbar}\{H,\r\}_-
+\L_{a,\kappa}\r\nn
&+\L_{\GX,\gamma_{\t{X}}}\r+\L_{\GD,\gamma_{\t{D}}}\r
\end{align}
with the commutator $\{A,B\}_-$ of operators $A$ and $B$.
We treat the phonon Hamiltonian in a numerically exact way based on a quasi-adiabatic path-integral (QUAPI) formalism.
\cite{Makri1995a,Makri1995b,Vagov2011,Barth2016,Cygorek2017}
By numerically exact we denote a solution that depends only on the time discretization and the memory length as the sole convergence parameters.
Beyond these two convergence parameters, no approximations enter the solution for the QD--cavity dynamics.

Physically, the phonon influence is fully captured by the phonon spectral density $J(\w)=\sum_{\q} \vert\g_{\q}\vert^2 \delta(\w-\w_{\q})$.
Assuming harmonic confinement and a linear dispersion $\w_{\q}=c_s|\q|$ with sound velocity $c_s$, the spectral density reads
\begin{align}
J(\w)=\frac{\w^3}{4\pi^2\r_D\hbar c_s^5}
\left(D_e e^{-\w^2a_e^2/(4c_s^2)}-D_h e^{-\w^2a_h^2/(4c_s^2)}\right)^2\, ,
\end{align}
where we have considered deformation potential coupling which is usually the dominant coupling mechanism.\cite{Krummheuer2002}
Here, $\r_D$ is the density of the material, $D_e$ ($D_h$) the electron (hole) deformation potential, and $a_e$ ($a_h$) the electron (hole) confinement radius, listed in Table~\ref{tab:par}.

\section{Buffering protocol}
\label{sec:buffering}

\subsection{General idea}
\label{subsec:idea}

\begin{figure}[t]
\centering
\includegraphics[width=\columnwidth]{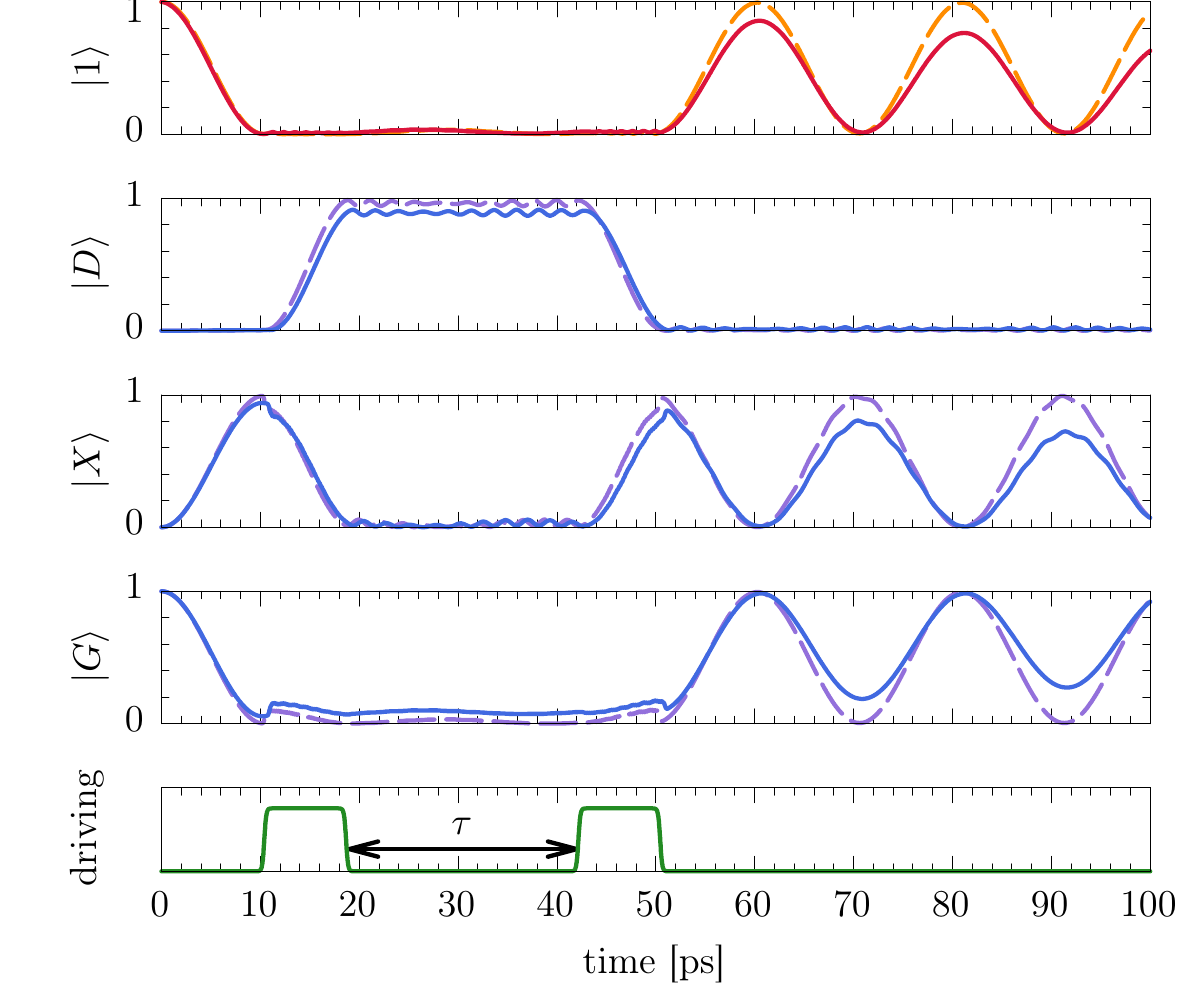}
\caption{A single cavity photon is stored in the dark exciton state of the QD using a first writing AC-Stark pulse (bottom panel).
The occupations of the ground state $\G$, the bright exciton $\X$, the dark exciton $\D$, and the $1$-photon Fock state $|1\>$ are depicted:
ideal case without phonons and losses (dashed lines);
including radiative and cavity loss effects (solid lines).
A second readout AC-Stark pulse retrieves the single photon.
The time between the pulses is the buffer time $\tau$.
}
\label{fig:single_dynamics}
\end{figure}

\begin{figure}[t]
\centering
\includegraphics[width=\columnwidth]{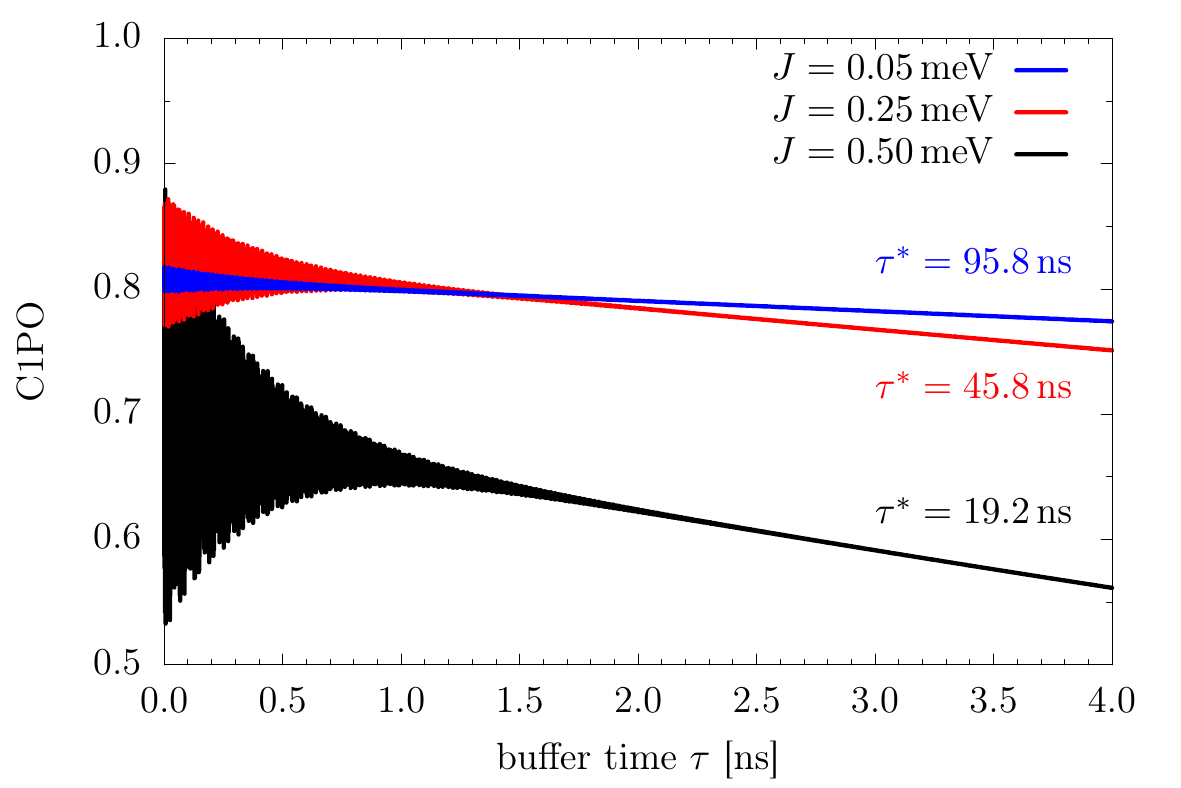}
\caption{The maximum captured $1$-photon occupation (C1PO) after the second AC-Stark pulse as a function of the buffering time $\tau$, i.e., the delay time between the two pulses.
The dependence is shown for different coupling strengths $J$ between the bright and the dark state.
Labels indicate the decay time $\tau^*$ extracted from fitting an exponential function to the corresponding curve (cf., main text for detailed explanation).
}
\label{fig:tau_J}
\end{figure}

We propose a protocol to buffer a single cavity photon deterministically using the $\Lambda$-type three-level system described in Section~\ref{sec:model}.
Initially, we assume the QD to be in its ground state $\G$ and one photon to be present in the cavity, i.e., the initial state of the QD--cavity dynamics is $|G,1\>$ (cf., \textbf{Figure~\ref{fig:single_dynamics}}), where we have introduced the notation $|\chi,n\>$ for the QD--cavity product space with $\chi\in\{G,X,D\}$ and the photon number $n$.
Due to the QD--cavity coupling, the cavity photon is absorbed into the bright state $\X$ after half a coherent Rabi oscillation.
Then, the dark state $\D$ is prepared using a recently proposed protocol relying on the optical Stark shift.\cite{Neumann2021}
The transfer of the excitation from the bright to the dark state is triggered deterministically using an off-resonant AC-Stark pulse, which shifts the bright state energy such that the bright and dark states are effectively in resonance.
The pulse duration is chosen such that exactly half a Rabi oscillation between the two exciton states is possible.

To present a physically clear picture of the processes involved in the buffering scheme, we use rectangular pulses with smoothed edges for the AC-Stark pulse envelopes, following Ref.~\onlinecite{Neumann2021}
\begin{align}
\label{eq:ACS_Pulse}
f_{\t{ACS}}(t)=\frac{f_0}{\left(1+e^{-\a(t-t_{\t{on}})}\right)\left(1+e^{-\a(t_{\t{ACS}}-(t-t_{\t{on}}))}\right)}\, .
\end{align}
Here, $\a$ determines the rise time of the pulse, which we set to $10\,$ps$^{-1}$, $t_{\t{on}}$ is the switch-on time, and $t_{\t{ACS}}$ the pulse duration.
The pulse amplitude $f_0$ is determined by the effective dark-bright splitting $\delta_{\t{eff}}$, which needs to be bridged, and the pulse duration $t_{\t{ACS}}$ by the oscillation frequency $J$.

During the pulse, when the amplitude is essentially $f_0$, the induced optical Stark shift is
\begin{align}
\Delta E_{\t{Stark}}=\frac{\hbar}{2}\left(\sqrt{\Delta\w_{\t{AX}}^2+f_0^2}-\Delta\w_{\t{AX}}\right)
\end{align}
for $\Delta\w_{\t{AX}}>0$.\cite{Neumann2021}
By setting $\Delta E_{\t{Stark}}=\delta_{\t{eff}}$, the pulse amplitude necessary to bridge the dark-bright splitting is determined to be
\begin{align}
f_0=\sqrt{\left(2\frac{\delta_{\t{eff}}}{\hbar}+\Delta\w_{\t{AX}}\right)^2-\Delta\w_{\t{AX}}^2}\, .
\end{align}

The length $t_{\t{ACS}}$ of the pulse has to correspond to half a Rabi oscillation between the two exciton states mediated by the spin-flip coupling\cite{Neumann2021}
\begin{align}
t_{\t{ACS}}=\frac{2\pi\hbar}{2\sqrt{J^2+\left(\delta_{\t{eff}}-\Delta E_{\t{Stark}}\right)^2}}\, ,
\end{align}
which simplifies to $t_{\t{ACS}}=\pi\hbar/J$ for $\Delta E_{\t{Stark}}=\delta_{\t{eff}}$.

This leaves only the detuning with respect to the bright exciton frequency $\Delta\w_{\t{AX}}$ as a free parameter.
For a dark-bright splitting in the order of a meV, it has been shown that a detuning of $\Delta\w_{\t{AX}}=15\,$meV is favorable for the transfer of the excitation from the bright to the dark state.\cite{Neumann2021}

In an ideal system without losses and decoherence, the excitation is expected to stay in the dark state indefinitely.
When including loss effects, it is important to note the different orders of magnitude of the dark and bright exciton decay rates.
Since the dark state is not optically active, it is a metastable state.
This is reflected in its decay rate $\gamma_{\t{D}}$ being about two orders of magnitude smaller than the radiative decay rate $\gamma_{\t{X}}$ of the bright state (cf., Table~\ref{tab:par}).
Therefore, the dark state is a good candidate for storing the photon in a realistic, lossy system.
The release of the photon is facilitated by the reverse process with a second AC-Stark pulse.

The time evolution of this protocol in the ideal case (without taking phenomenological losses or phonons into account) is presented in Figure~\ref{fig:single_dynamics} (dashed lines), which shows the occupation of the three states together with the occupation of the $1$-photon state and the applied laser pulses as functions of time.
The dynamics behaves as predicted by the \textit{writing} scheme described above.
Indeed, the occupation of the dark state after the first writing pulse is close to unity.
Small-amplitude oscillations appear due to the residual coupling to the bright state, which depend both on the coupling $J$ and the splitting $\delta_{\t{eff}}$ between the bright and the dark state.

To release the photon after the buffering time $\tau$ ($23.5\,$ps in the example shown in Figure~\ref{fig:single_dynamics}), a second \textit{readout} AC-Stark pulse is required (cf., bottom panel of Figure~\ref{fig:single_dynamics}).
When the excitation is transferred back to the single-photon state, Rabi oscillations between the cavity mode and the bright exction are observed.
These oscillations are undamped in the ideal case, where no phonon coupling and no phenomenological loss processes are considered (cf., dashed lines in Figure~\ref{fig:single_dynamics}).
The maximum occupation of the $1$-photon state $|1\>$ is $99.95\,\%$, implying a close to perfect writing and readout of the buffered single photon in the ideal case.



\section{Storage performance}
\label{sec:performance}


The key quantity of interest in a buffering scheme is the retrievable percentage of the stored photon after the buffering time $\tau$.
Therefore, we here discuss the dependence of this captured $1$-photon occupation on the buffering time $\tau$ and various system parameters, including the dark-bright coupling $J$, the splitting $\delta_{\t{eff}}$, and the temperature $T$.
An example of the influence of phenomenological losses on the buffering scheme is shown in Figure~\ref{fig:single_dynamics}(b) (solid lines).
We find that the scheme is degraded and here we quantify the amount of storage which is still achievable.
Since damped Rabi oscillations between the bright exciton and the cavity occur in the protocol proposed in Section~\ref{sec:buffering} after the readout AC-Stark pulse when including phenomenological losses, we take the maximum captured $1$-photon occupation (C1PO) after the readout pulse as a measure of the retrievable percentage of the stored photon.

\subsection{Influence of the dark-bright coupling $J$}
\label{subsubsec:J}

\textbf{Figure~\ref{fig:tau_J}} shows the dependence of the C1PO on the buffering time $\tau$, i.e., the delay time between the two AC-Stark pulses.
These calculations are performed considering phenomenological losses, i.e., bright and dark exciton decay and cavity losses, but without taking phonons into account.
The red line corresponds to the coupling $J=0.25\,$meV.
After initial oscillations, the C1PO decreases exponentially.
The oscillations are a direct consequence of the dark-bright coupling with strength $J$, which is off-resonant due to the dark-bright splitting $\delta_{\t{eff}}$.
This off-resonance leads to low-amplitude high-frequency oscillations of the dark state occupation in between the writing and readout pulses (cf., Figure~\ref{fig:single_dynamics}).
When the second AC-Stark pulse arrives during a minimum of this oscillation, the corresponding value of the C1PO also becomes minimal.

The damping of these oscillations and the subsequent exponential decay shown in Figure~\ref{fig:tau_J} is a result of the decay of the dark state $\D$.
While its intrinsic decay rate $\gamma_{\t{D}}$ corresponds to a lifetime of $100\,$ns, the overall effective decay of the dark state depends not only on this decay time, but also on the mixing between the dark and the bright state due to their residual off-resonant coupling.
Since the admixture of the bright state decays on the much shorter time scale of $\gamma_{\t{X}}^{-1}=0.4\,$ns, the combined effective decay time of the dark state becomes much faster (cf., analytical discussion in Appendix~\ref{app:derivation}).

Fitting an exponential function of the form
\begin{align}
\label{eq:fit}
\t{C1PO}_{\t{fit}}(\tau)=c\,e^{-\tau/\tau^*}
\end{align}
to C1PO($\tau$) in Figure~\ref{fig:tau_J} using the scaling constant $c$ and the decay time $\tau^*$ as free parameters, one obtains a decay time of $\tau^*=45.8\,$ns (cf., red line in Figure~\ref{fig:tau_J}).
This value agrees well with the effective decay time of the dark state as derived in the appendix [cf., also Eq.~\eqref{eq:analytic}], corroborating the conclusion that the storage performance between writing and readout only depends on the decay of the dark state $\D$.

Increasing $J$ while keeping all the other parameters fixed, yields a shorter decay time $\tau^*$ (cf., black line in Figure~\ref{fig:tau_J}) and thus a worse performance of the storage protocol.
The reason is the increased oscillation amplitude between $\D$ and $\X$ stemming from the larger dark-bright coupling.
Therefore, the interaction with the faster-decaying bright state is more effective.
The higher oscillation amplitude is reflected in the larger initial oscillation amplitude of C1PO.
The reverse argument holds for a smaller coupling strength $J$ and indeed for $J=0.05\,$meV (cf., blue line in Figure~\ref{fig:tau_J}) $\tau^*=95.8\,$ns is already close to the lifetime of $100\,$ns of the dark exciton without exchange coupling to the bright exciton.

Overall, decay times on the order of a few tens of ns suggest a high storage performance.
In comparison, a single photon inside a high-Q cavity with a quality factor of $2.68\cdot10^5$,\cite{Schneider2016} corresponding to our value of $\kappa$ (cf., Table~\ref{tab:par}), has a decay time of $\kappa^{-1}=118\,$ps.
Therefore, the buffering protocol presented here facilitates a storage time roughly two orders of magnitude longer.

Note that changing the dark-bright coupling $J$ experimentally means that the location of the Mn atom $\bm{r}_{\t{Mn}}$ needs to be changed, thus requiring different QD samples.
Changing the Mn position also changes the shifts induced by the carrier-Mn Ising terms.
Therefore, an additional magnetic field in Faraday configuration would be necessary to keep $\delta_{\t{eff}}$ constant [cf., Eq.~\eqref{eq:delta_eff}].

\subsection{Influence of the dark-bright splitting $\delta_{\t{eff}}$}
\label{subsubsec:delta_eff}

\begin{figure*}[t]
\begin{minipage}[t]{0.49\textwidth}
 \centering
 \includegraphics[width=\textwidth]{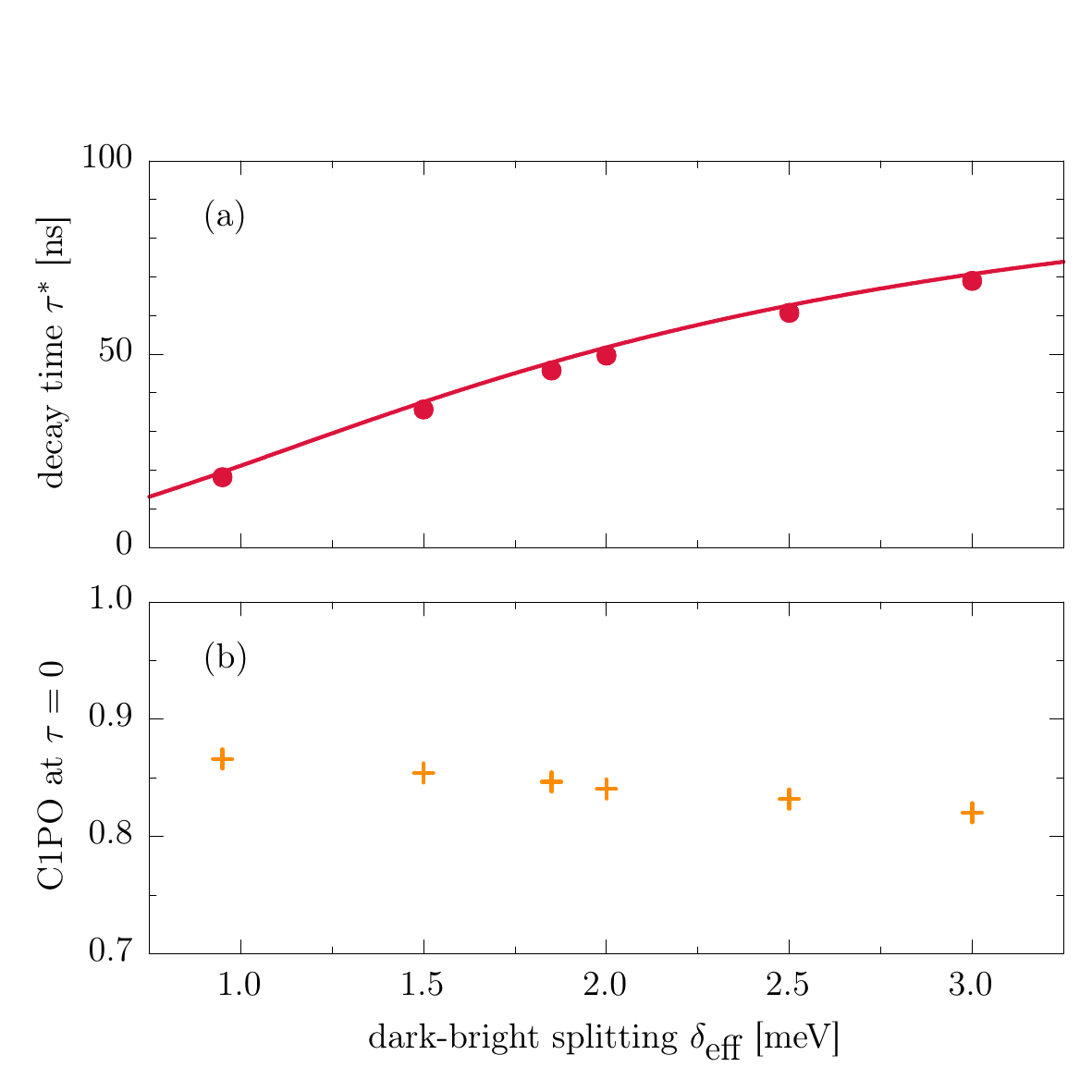}
\end{minipage}
\begin{minipage}[t]{0.49\textwidth}
 \centering	
 \includegraphics[width=\textwidth]{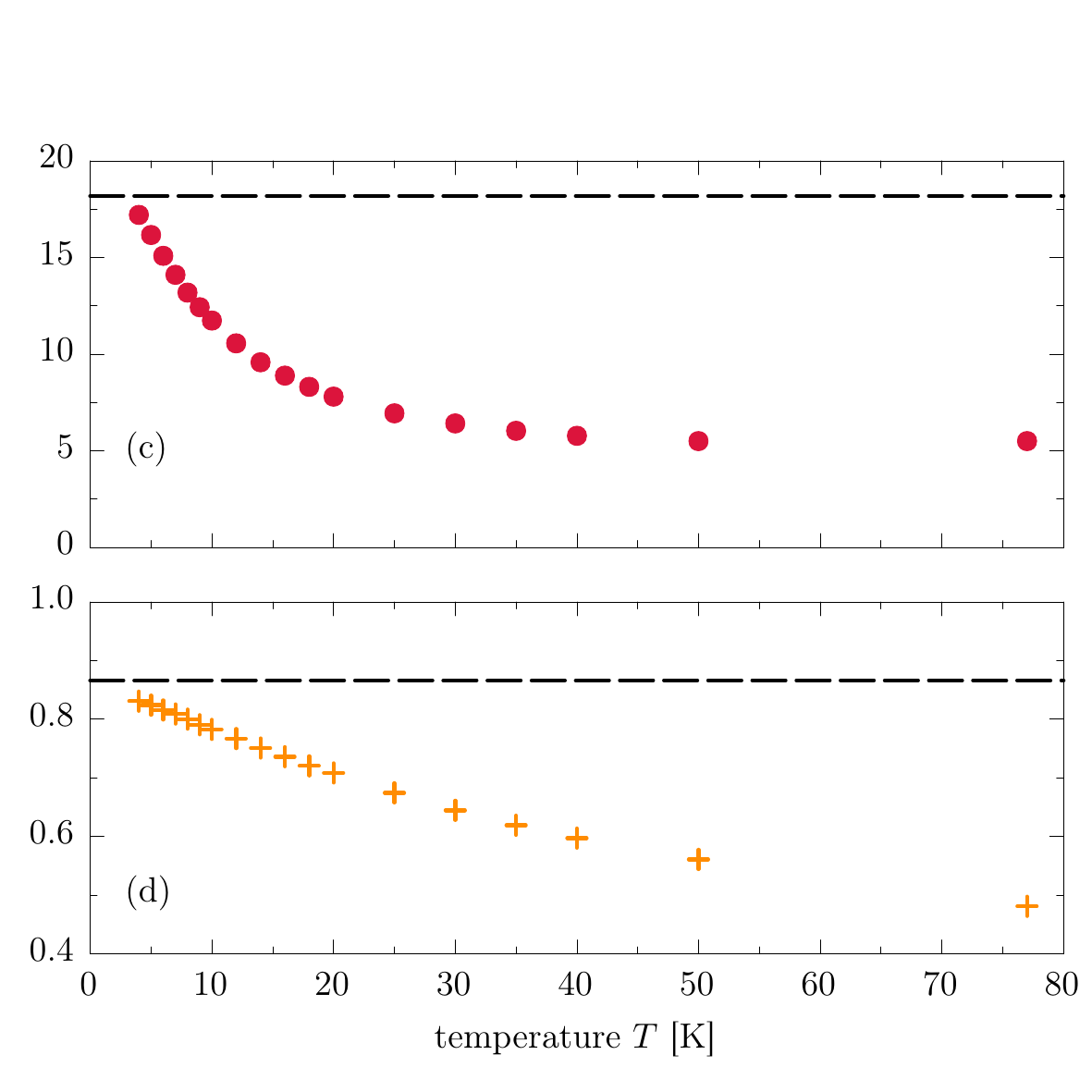}
	\end{minipage}
\caption{The dependencies of the decay time $\tau^*$ and the C1PO at $\tau=0$ on the effective splitting $\delta_{\t{eff}}$ between the dark and bright state [panels (a) and (b)] without taking phonons into account and on the temperature $T$ [panels (c) and (d)] including phonon effects.
The red solid line in panel (a) is the approximate analytical prediction from Equation~\eqref{eq:analytic}.
For the temperature dependent study, the splitting is set to $\delta_{\t{eff}}=\delta_{\t{XD}}=0.95\,$meV.
}
\label{fig:dXD_T_analysis}
\end{figure*}

In the previous section, it became clear that the main loss channel during the storage time is the effective decay of the dark state.
This in turn depends on the residual coupling to the bright state, which is determined by $J$ and $\delta_{\t{eff}}$.
In this section, we analyze the influence of the latter by repeating the calculations of C1PO$(\tau)$ by varying $\delta_{\t{eff}}$ and fitting Eq.~\eqref{eq:fit} to the resulting curves, in analogy to Figure~\ref{fig:tau_J}.
We keep the dark-bright coupling fixed at $J=0.25\,$meV.

It is instructive to investigate the dependence of the decay time $\tau^*$ on the effective dark-bright splitting $\delta_{\t{eff}} $ shown in \textbf{Figure~\ref{fig:dXD_T_analysis}}(a).
We note that while this analysis is a theoretical parameter study and in principle would require to analyze several QDs, it could also be performed on the same QD by applying an external magnetic field in Faraday configuration with magnitude $B_z$ to control $\delta_{\t{eff}}$ [cf., Eq.~\eqref{eq:delta_eff}]. 

We vary the splitting around the value of $\delta_{\t{eff}}=1.85\,$meV\cite{Furdyna1988,Besombes2002}.
The corresponding decay time $\tau^*$ is the same as the one obtained from the red line in Figure~\ref{fig:tau_J}. 
At constant $J$, a higher splitting means that the residual coupling of the dark to the bright state is weaker.
This closes the corresponding radiative loss channel more and more, such that $\tau^*$ converges to the intrinsic decay time of the dark state $\gamma_{\t{D}}^{-1}=100\,$ns.
The opposite argument holds for smaller splittings.
Without any splitting, the two exciton states would perform coherent full-amplitude Rabi oscillations with a frequency corresponding to $J$, such that the radiative decay channel would diminish the storage performance maximally.

Assuming $g\ll\delta_{\t{eff}}$ and $J\ll\delta_{\t{eff}}$, which holds well for the parameters considered in Figure~\ref{fig:dXD_T_analysis}(a), an analytical approximation of the following form can be derived (for a detailed derivation, see Appendix~\ref{app:derivation}):
\begin{align}
\label{eq:analytic}
\tau^*(\delta_{\t{eff}})=\left[\left(\frac{J}{2\delta_{\t{eff}}}\right)^2 \left(\gamma_{\t{X}}-\gamma_{\t{D}}\right)+\gamma_{\t{D}}\right]^{-1}\, .
\end{align}
This function is plotted in \textbf{Figure~\ref{fig:dXD_T_analysis}}(a) as a red solid line, reproducing the numerically obtained data well.

To analyze the performance of the writing and reading process separately from the losses during storage, we take the C1PO for $\tau=0$ as a measure, i.e., the writing and readout pulses merge to a single pulse of length $2t_{\t{ACS}}$.
This value indicates, what percentage of the initially present photon can be retrieved after writing it to the dark state and immediately reading it out again.
Note that due to the initial oscillations of C1PO$(\tau)$ (cf., data shown in Figure~\ref{fig:tau_J}), the fit parameter $c$ in Eq.~\eqref{eq:fit} does not necessarily correspond to the value of C1PO$(\tau=0)$.

The results are shown in Figure~\ref{fig:dXD_T_analysis}(b).
Overall, the losses during writing and readout are restricted to values between $10\,\%$ and $20\,\%$, originating from the loss rates $\gamma_\t{X}$ and $\kappa$.

\subsection{Temperature dependence}
\label{subsubsec:T_dependence}

Including the coupling of both the bright and the dark exciton states to LA phonons as described by Equation~\eqref{eq:H_Ph} leads to a faster decay of the initial oscillations of C1PO$(\tau)$ and a faster subsequent exponential decay.
We perform this analysis for a dark-bright splitting of $\delta_{\t{eff}}=\delta_{\t{XD}}=0.95\,$meV, i.e., for the intrinsic splitting due to the electron-hole exchange, with a coupling of $J=0.25\,$meV.

\begin{figure*}[t]
\begin{minipage}[t]{0.49\textwidth}
\centering
\includegraphics[width=\textwidth]{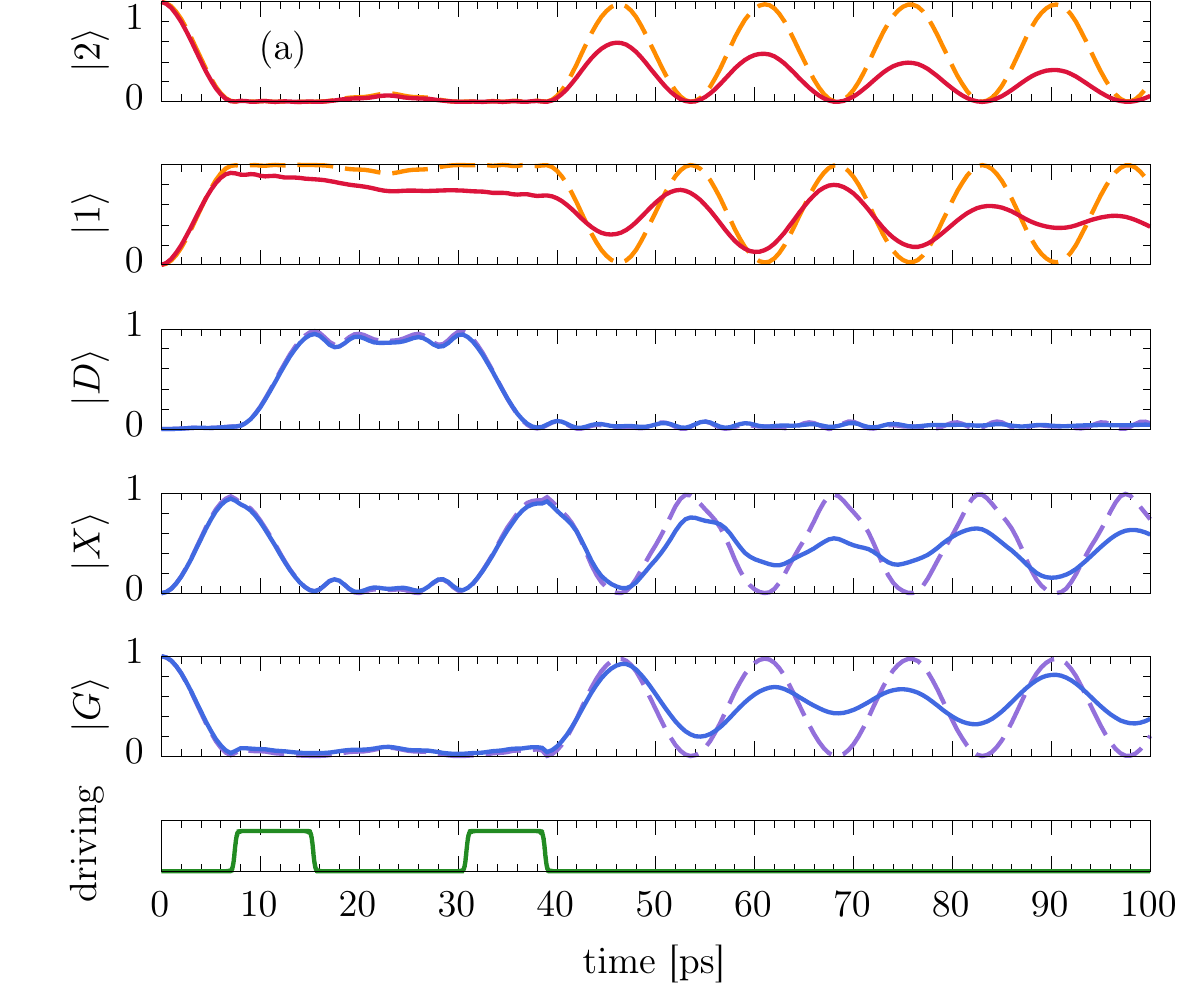}
\end{minipage}
\begin{minipage}[t]{0.49\textwidth}
\centering
\includegraphics[width=\textwidth]{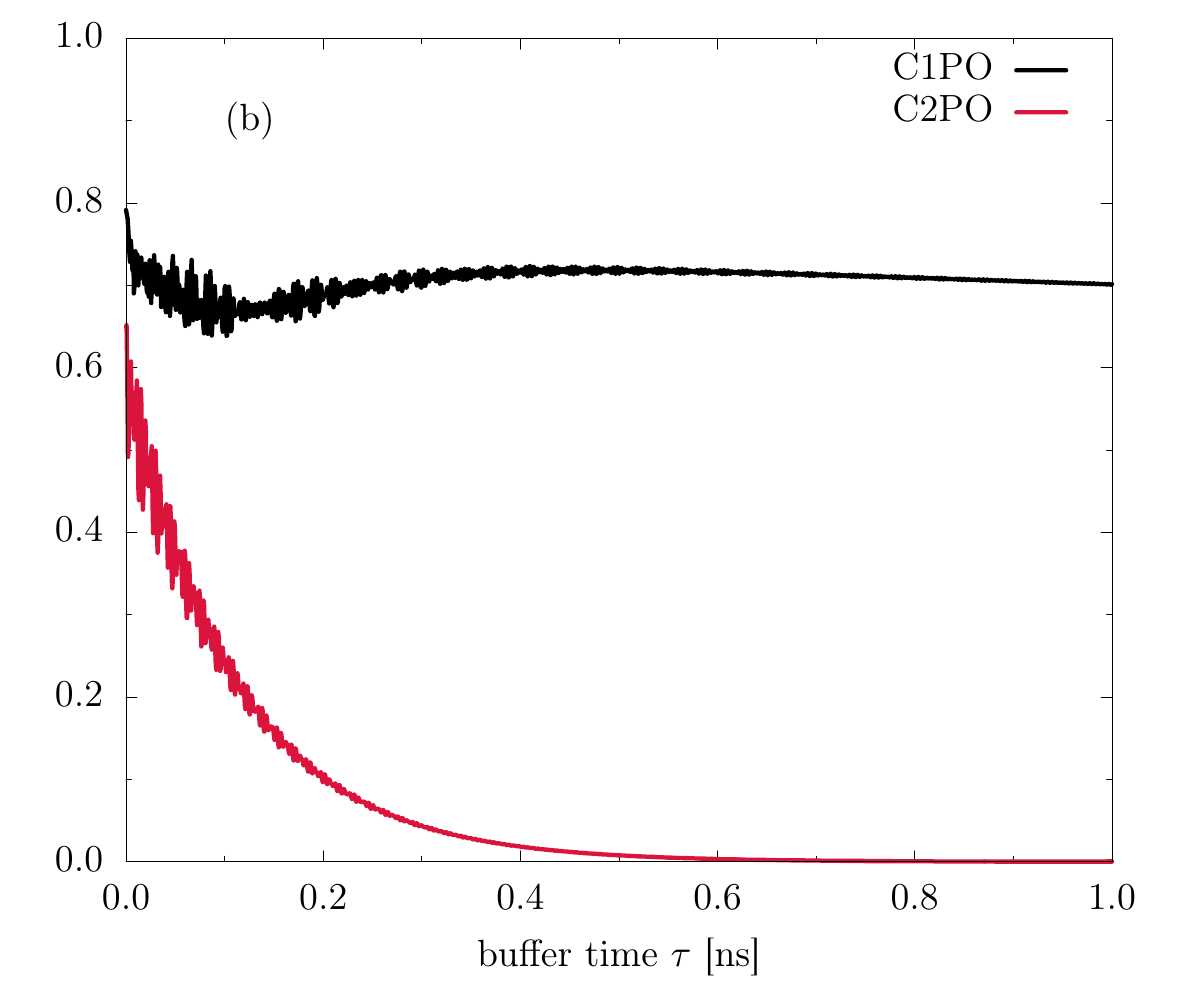}
\end{minipage}
\caption{(a) Same protocol as in Figure~\ref{fig:single_dynamics}, but for an initial QD--cavity state of $|G,2\>$.
While one of the photons is stored in the dark state, the remaining one can leave the cavity, when cavity losses are taken into account.
Ideal case without phonons and losses (dashed lines);
including radiative and cavity loss effects (solid lines)
(b) The maximum captured photon occupation after the second AC-Stark pulse as a function of the buffering time $\tau$.
Here, both phenomenological loss effects and the phonon influence at $T=4\,$K are taken into account.
}
\label{fig:N_2}
\end{figure*}

The resulting decay times $\tau^*$ and values of C1PO$(\tau=0)$ are shown in Figure~\ref{fig:dXD_T_analysis}(c) and (d), respectively.
The phonon-free results are marked by dashed black lines.
At $T=4\,$K, the decay time is close to its phonon-free counterpart.
With rising temperature, though, the decay times become drastically shorter.
At $T=77\,$K, it is only roughly a quarter of the phonon-free value.
The reason is the asymmetry of phonon absorption and emission at low temperatures that vanishes at higher temperatures.
During storage, the state $|D,0\>$ is mostly occupied.
The state $|G,1\>$ lies $\delta_{\t{eff}}=0.95\,$meV above it and thus cannot be reached by phonon emission, which is predominant at low temperatures.
In contrast, at higher temperatures, $\delta_{\t{eff}}$ can be bridged by phonon absorption.
Thus, an additional decay channel of the dark state $|D,0\>$ opens during storage.
Reducing the residual coupling during storage by means of smaller $J$ or a $\delta_{\t{eff}}$ much larger than the maximum of the phonon spectral density should therefore weaken the phonon influence, too.

The losses due to writing and readout are also hardly influenced at low temperatures, while they become stronger with rising $T$.
This means that the preparation of the dark state during writing is already incomplete.
The reason lies in the fact that the phonon interaction dampens the Rabi oscillations between the bright state and the cavity to an extent that already the transfer from the single-photon state to the bright exciton (before the writing pulse) is incomplete.

\section{Storage of a single photon out of the state $|n\>$ with $n>1$}
\label{subsec:N_2}

We have demonstrated the buffering capacity of our protocol concerning a single-photon state.
Now, the question arises how it performs, when higher-order Fock states are present in the cavity.
To this end, we consider the state $|G,2\>$ as the initial value of the QD--cavity system and buffer one of the two photons present in the cavity using the presented protocol.
The occupation dynamics is shown in \textbf{Figure~\ref{fig:N_2}}(a) for a fixed $\tau=15.5\,$ps.
The analysis is performed for $J=0.25\,$meV and $\delta_{\t{eff}}=\delta_{\t{XD}}=0.95\,$meV as before.
Dashed lines show the ideal case, while solid lines depict the case including phenomenological losses.

We consider both the C1PO and the captured $2$-photon occupation (C2PO) after the buffering time $\tau$ in Figure~\ref{fig:N_2}(b) (black and red lines, respectively).
All loss processes and the phonon influence at $T=4\,$K are taken into account in these results.
The C2PO decays exponentially.
The rate corresponds exactly to the cavity loss rate $\kappa$.
Since one of the two initially present photons is stored in the dark state, the remaining single photon can leave the cavity via the cavity loss channel.
Retrieving the other photon from the dark state and recombining it with the remaining one to yield the initial Fock state $|2\>$ is only possible, when the remaining one has not left the cavity yet.

Nonetheless, the effective buffering of the $2$-photon Fock state outperforms the case, where the state $|2\>$ decays without using a storage scheme.
The reason is the fact that the Fock state $|n\>$ decays with an effective rate of $n\kappa$.
Therefore, our single-photon buffering protocol can reduce this effective rate to $(n-1)\kappa$, as shown here for the case $n=2$.
Meanwhile, the dependence of the C1PO on the buffering time corresponds again to the effective lifetime of the dark state of about $\sim20\,$ns.
Interestingly, the dependence of the C1PO on $\tau$ is, even when disregarding the high-frequency oscillations in the beginning, nonmonotonous.
The reason is the photon that remains in the cavity:
the Rabi frequency of the oscillations between the bright state and the cavity depends on the number of photons present in the cavity.
Since the frequencies for the different photon numbers are incommensurable, changes in the amplitude and therefore the nonmonotonicity of the C1PO are the consequence.

\section{Towards experimental realization}
\label{sec:experimental}

To present a clear and well understandable physical picture of the buffering scheme, we used rectangular pulses with smoothed edges as model AC-Stark pulses [cf., Eq.~\eqref{eq:ACS_Pulse}].
While such pulses can be generated using fast electro-optical modulators to cut the desired envelopes out of a continuous wave laser,\cite{Neumann2021} the rise time of $1/\a=0.1\,$ps assumed in Section~\ref{subsec:idea} in combination with the pulse length necessary for the protocol is out of reach with current state-of-the-art equipment.\cite{Neumann2021}
Experimentally, it is a far lesser challenge to use pulses with Gaussian envelopes.

The AC-Stark pulses are needed for the excitation transfer from the bright to the dark state for the writing and vice versa for the readout procedure.
Therefore, we compare the storage capacity of differently shaped pulses by using the maximum occupation of the dark state $\D$ after the first (writing) pulse (cf., Figure~\ref{fig:single_dynamics}) as a target quantity in the following.
Note that any losses experienced during writing occur again at readout, thus influencing the C1PO two times.
Nonetheless, the pulse shape should not have any influence on the decay time $\tau^*$ during storage, since there are no pulses in the time interval between writing and readout.

Using Gaussian pulses of the form
\begin{align}
f_{\t{ACS}}(t)=\frac{\Theta}{\sqrt{2\pi}\,\s}e^{-\frac{(t-t_0)^2}{2\s^2}}\, ,
\end{align}
three parameters have to be determined:
the pulse area $\Theta$, the standard deviation $\s$, which is connected to the full width at half maximum via $\t{FWHM}=2\sqrt{2\ln{2}}\,\s$, and the time $t_0$, where the maximum of the pulse occurs.
While the three parameters $f_0$, $t_{\t{ACS}}$, and $t_{\t{on}}$ can be determined from analytical considerations for rectangular pulses from Ref.~\onlinecite{Neumann2021}, predicting an optimal set of Gaussian pulse parameters is not straightforward.
Therefore, we numerically search for the maximum occupation of $\D$ in the parameter space spanned by $\Theta$, $\s$, and $t_0$ by discretizing all three parameters.
At $T=4\,$K for example, this optimum is given by $\Theta=33.77\pi$, $\t{FWHM}=7.14\,$ps, and $t_0=15.01\,$ps.
We perform this optimization for the parameters used in the last two sections, namely $J=0.25\,$meV and $\delta_{\t{eff}}=\delta_{\t{XD}}=0.95\,$meV.

\begin{figure}[t]
\centering
\includegraphics[width=\columnwidth]{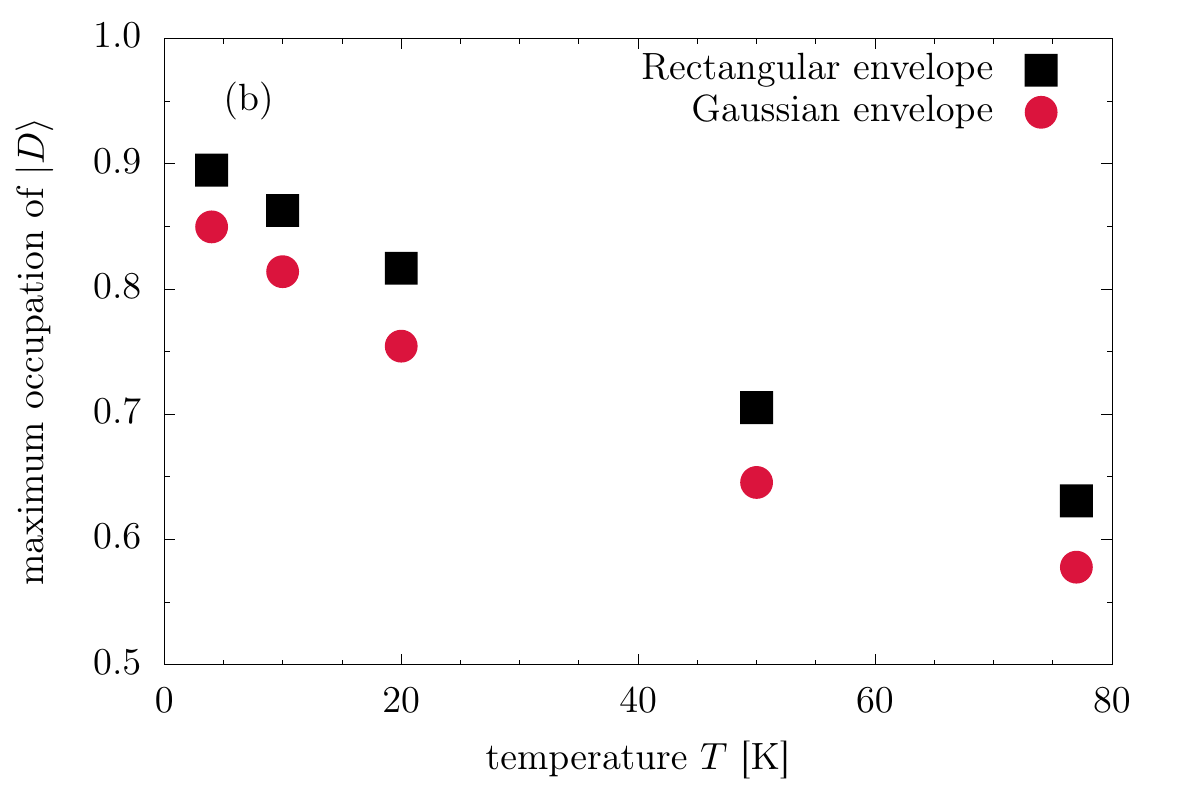}
\caption{The maximum occupation of the dark exciton $\D$ after the storage pulse depending on temperature.
Pulses with rectangular envelope are compared with Gaussian pulses.
}
\label{fig:dark_temperature}
\end{figure}

\textbf{Figure~\ref{fig:dark_temperature}} shows the results depending on the temperature $T$.
Although the rectangular pulses consistently outperform the Gaussian ones, the loss in occupation due to the experimentally easier to implement Gaussian shape is only around $5$ percentage points for all considered temperatures.
Therefore, the presented buffering protocol also works with Gaussian instead of rectangular pulses.
This provides a path to an experimental realization in the near future.


Finally, let us comment on the usage of a semimagnetic QD for this protocol.
The dark-bright coupling provided by the Mn atom is crucial for the operation of the protocol.
The advantage of the Mn doping is that the dark-bright interaction is provided by an intrinsic degree of freedom of the QD, thus pointing the way towards creating compact on-chip devices.
Nonetheless, this coupling could also be provided by an additional external magnetic field in Voigt configuration without using the Mn atom as a mediator.\cite{Neumann2021}
Therefore, our proposed storage protocol should in principle also work in nonmagnetic QDs, but the controlled application of a magnetic field only to the buffering device could be obstructive in miniaturized circuits containing several components.

\section{Conclusion} 
\label{sec:conclusion}

We proposed a protocol to deterministically write and read a single photon in a QD--cavity system.
Assuming a CdTe QD isoelectrically doped with a single Mn ion yields a $\Lambda$-type three-level system consisting of a ground state and two exciton states, one optically active bright state and one that is dipole dark.
The storage protocol relies on a coherent transfer of the photon occupation to the bright exciton due to Rabi oscillations.
Then, an AC-Stark pulse shifts the bright state to be in resonance with the dark exciton.
A coherent excitation transfer during the length of the pulse prepares the dark state, which due to its optical inactivity is a metastable state with long lifetime.
The readout procedure is exactly the reverse process.


We analyzed the influence of the dark-bright coupling strength $J$ and the effective dark-bright splitting $\delta_{\t{eff}}$ on the performance of the protocol as well as its dependence on temperature.
During storage in the dark state, its residual coupling to the bright state and thus to faster loss channels is controlled by $J$ and $\delta_{\t{eff}}$.
Reducing this residual coupling by decreasing $J$ or increasing $\delta_{\t{eff}}$ leads to a better overall performance of the buffering scheme.
At rising temperatures, the phonon environment acts on the coupling between the bright state and the cavity.
Thus, an additional loss channel during storage has to be considered, which again can be influenced by adjusting the residual coupling of the dark to the bright exciton.
Furthermore, phonons have a rather strong influence on the writing and readout procedure.
At high enough temperatures, already the transfer of the photon to the bright exciton before writing becomes incomplete.

Nonetheless, for all considered parameter sets the overall storage time as measured by $\tau^*$ ranges from a few to tens of ns.
Thus, it is two orders of magnitude longer than the lifetime of a photon in a high-Q cavity with a quality factor of $2.68\cdot10^5$.\cite{Schneider2016}

Furthermore, we have shown that the proposed scheme can store a single photon out of a higher-order Fock state $|n\>$ with $n>1$.
Thus, the lifetime of the state $|n\>$, which is $(n\kappa)^{-1}$ in a cavity with loss rate $\kappa$, can be extended to $[(n-1)\kappa]^{-1}$ (for $\gamma_{\t{D}}\ll\kappa$).

Finally, we discussed the possibility of using Gaussian pulses for the buffering protocol instead of rectangular ones, which are experimentally out of reach with current equipment for the pulse characteristics needed for the protocol.
For optimal pulse parameters, Gaussian pulses can be used successfully.
Pulses of rectangular shape are only $\sim5$ percentage points better concerning the dark state occupation after the writing procedure.

Thus, we expect the proposed scheme to be realizable with state-of-the-art equipment.
After QDs have long been discussed as on-demand single-photon sources, this work paves the way for them to also be used as storage components. A main advantage of using magnetically doped QDs is that no external magnetic field is necessary. Such a QD buffering device for single photons could serve as a building block in more complex QD quantum information processing devices.

\renewcommand{\theequation}{A\arabic{equation}}
\renewcommand{\thesection}{A}
\renewcommand{\thesubsection}{A.\arabic{subsection}}
\setcounter{equation}{0}


\medskip
\textbf{Conflict of Interest} \par 
The authors declare no conflict of interest.

\medskip
\textbf{Acknowledgements} \par 
This work was funded by the Deutsche
Forschungsgemeinschaft (DFG, German Research Foundation) -- project No. 419036043.
M.Co. gratefully acknowledges support by the Studienstiftung des Deutschen Volkes.

\medskip
\appendix
\section{Derivation of the effective decay rate}
\label{app:derivation}
Here, we give a brief derivation of the analytic equation given for the effective decay rate in Eq.~\eqref{eq:analytic}. In the basis $\left\lbrace \KET{D,0}, \KET{X,0}, \KET{G,1}\right\rbrace$, the Hamiltonian $H$, describing the coherent part of the system dynamics during the storage time, is given by:
\begin{equation}
H =
\begin{pmatrix}
-\deff & -\frac{1}{2}J & 0 \\
-\frac{1}{2}J & 0 & \hbar g \\
0 & \hbar g & 0 \\
\end{pmatrix}.
\end{equation}
The formal expression for any of the three eigenstates $\KET{\lambda}$ of $H$ is
\begin{equation}
\label{eq:formal_lambda}
\KET{\lambda} = \mathcal{N}_\lambda \bigg\lbrace \KET{D,0} - \frac{J}{2E_\lambda \big[ 1-\big(\frac{\hbar g}{E_\lambda}\big)^2 \big]} \left[ \KET{X,0} + \frac{\hbar g}{E_\lambda} \KET{G,1} \right] \bigg\rbrace\, ,
\end{equation}
where $\mathcal{N}_\lambda$ is a normalization constant and $E_\lambda$ is the corresponding eigenenergy. In the situation considered here, both coupling constants of the model are much smaller than the effective dark-bright splitting, i.e. $J/(2\deff),\hbar g/\deff \ll 1$. Therefore the energetically lowest eigenstate $\KET{d}$ has an energy $E_d\sim -\deff$, which is on the order of the dark-bright splitting, and is thus associated with the dark exciton. Furthermore, the contribution of the state $\KET{G,1}$ to $\KET{d}$ is then on the order $J\hbar g/\deff^2$, cf., Eq.~\eqref{eq:formal_lambda}, and can be neglected. In other words, this means that the coupling to the cavity mode has only a negligible impact on this eigenstate and $\KET{d}$ can in good approximation be written as the energetically lower eigenstate of the upper left 2$\times$2-matrix of $H$:
\begin{subequations}
\begin{equation}
\KET{d} = c_{\t{D}} \KET{D,0} + c_{\t{X}} \KET{X,0}\, ;
\end{equation}
\begin{equation}
c_{\t{X}} = \frac{J}{\sqrt{J^2+\left( \deff +\sqrt{\deff^2+J^2} \right)^2 }} \hs c_{\t{D}} = \sqrt{1-c_{\t{X}}^2}\, ; 
\end{equation}
\begin{equation}
E_d = -\frac{1}{2}\left( \deff + \sqrt{\deff^2+J^2} \right)\, .
\end{equation}
\end{subequations}
Because $\KET{d}$ is strongly associated with the dark exciton state $\KET{D,0}$ the decay of the excitation during the storage time is determined by the effective decay rate $\gamma_\text{eff}$ of this eigenstate. Considering the loss processes via the three Lindblad superoperators $\LB{a}{\kappa}$, $\LB{\OP{G}{X}}{\gamma_\text{X}}$, and $\LB{\OP{G}{D}}{\gamma_\text{D}}$ leads to a contribution
\begin{equation}
\ddt \rho_d = - \left( c_{\t{X}}^2 \gamma_\text{X} + c_{\t{D}}^2 \gamma_\text{D} \right) \rho_d = -\gamma_\text{eff}\,\rho_d
\end{equation}
in the dynamical equation for the occupation $\rho_d = \BRA{d}\rho\KET{d}$ of the state $\KET{d}$. Keeping only terms up to the second order in the small parameter $J/(2\deff)$ yields
\begin{equation}
\gamma_\text{eff} \approx \left(\frac{J}{2\deff}\right)^2  \gamma_\text{X} + \left[1-\left(\frac{J}{2\deff}\right)^2\right]\gamma_\text{D}.
\end{equation}
Consequently, the corresponding decay time of the stored excitation is given by $\tau^\ast = \gamma_\text{eff}^{-1}$, which directly leads to the analytic expression given in the main text.

\bibliography{bib,Photon_Buffer}

\end{document}